\documentclass{aa}
\usepackage{graphicx}
\begin{document}

\title{Analysing the database for stars in open clusters
I. General methods and description of the data}
\author{J.-C.~Mermilliod\inst{1}, E.~Paunzen\inst{2,3}}

\mail{Jean-Claude.Mermilliod@obs.unige.ch}

\institute{Institut d'Astronomie de l'Universit{\'e} de Lausanne, 
		   CH-1290 Chavannes-des-Bois, Switzerland
\and       Institut f\"ur Astronomie der Universit\"at Wien,
           T\"urkenschanzstr. 17, A-1180 Wien, Austria
\and       Zentraler Informatikdienst der Universit\"at Wien,
           Universit\"atsstr. 7, A-1010 Wien, Austria}

\date{Received 7 May 2003/Accepted 25 June 2003}
\titlerunning{Analysing the database for stars in open clusters I.}{}

\abstract{We present an overview and statistical analysis of the data
included in WEBDA. This database includes valuable information such as coordinates, 
rectangular positions, proper motions, photometric as well as spectroscopic data, radial and
rotational velocities for objects of open clusters in our Milky Way. It also contains 
miscellaneous types of data like 
membership probabilities, orbital elements of spectroscopic binaries and periods of 
variability for different kinds of variable stars. Our final goal is to derive
astrophysical parameters (reddening, distance and age) of open clusters
based on the major photometric system which will be presented in a follow-up paper. 
For this purpose
we have chosen the Johnson $UBV$, Cousins $VRI$ and Str{\"o}mgren $uvby\beta$
photometric systems for a statistical analysis of published data sets included
in WEBDA. Our final list contains photographic, photoelectric and
CCD data for 469820 objects in 573 open clusters.
We have checked the internal (data sets within one photometric system and the same
detector technique) and external (different detector technique) accuracy and conclude
that more than 97\,\% of all investigated data exhibit a sufficient accuracy for our
analysis. The way of weighting and averaging the data is described. In addition,
we have compiled a list of deviating measurements which is available to the
community through WEBDA.
\keywords{Open clusters and associations: general -- catalogs}
}
\maketitle

\section{Introduction}

The study of open clusters is very important in several respects. It allow one
to estimate different important astrophysical parameters within individual
clusters as well as the study of the wider solar neighbourhood concerning
its structure.

For this purpose it is essential to have a homogeneous set of photometric and additional
(e.g. membership probability, proper motion) data for a statisticaly significant
number of open clusters.

One of the most compelling databases in this respect is the WEBDA interface which
has been developed at the Institute for Astronomy 
at the University of Lausanne, Switzerland. It offers astrometric data in the form 
of coordinates, rectangular positions, and some proper motions, photometric data 
in the major systems in which star clusters have been observed (e.g. Johnson-Cousins
$UBVRI$, Str{\"o}mgren $uvby\beta$ and Geneva 7-color), spectroscopic data, 
like spectral classification, radial 
velocities, rotational velocities. It also contains miscellaneous types of data like 
membership probabilities, orbital elements of spectroscopic binaries and periods of 
variability for different kinds of variable stars. Finally a whole set of bibliographic 
references allows the community to locate the relevant publications for each
individual cluster easily.

In this first paper we present the compilation of 573
open clusters for which photometric measurements are available within WEBDA. The statistical
methods used to derived weighted means are described. Lists with objects showing
deviating photometric measurements within one system and/or different observing 
techniques (photographic, photoelectric and CCD) were generated. We discuss the internal
and external measurement accuracies based on a statistically significant sample of
independent sources from the literature. 

Our final goal of paper II will be the determination of the ages, distances
and reddening for the presented open clusters using the newest isochrones. This 
analysis will include the Johnson-Cousins $UBVRI$ and Str{\"o}mgren $uvby\beta$ photometric
systems and should supersede the work of 
Janes \& Adler (1982) who presented a compilation of 434 open cluster of our Milky Way
for which they summarized the reddening, ages and distances from 610 references in 
order to analyse the galactic structure. Their compilation was highly nonuniform
since they made no attempt to redetermine the appropriate astrophysical parameters.

\begin{table}[t]
\begin{center}
\caption{Excerpt of the content of WEBDA from the 8th of April
2003.}
\label{content}
\begin{tabular}{lrrr}
\hline
Data description & Clusters & Measurements & Stars \\
\hline
{\bf Fundamental} \\ 
Identifications & 403 & 12079 & 12055 \\ 
Transit Tables & 315 & & 349171 \\ 
Coordinates J2000 & 408 & 110385 & 109256 \\ 
Coordinates B1950 & 480 & 143775 & 134028 \\ 
Positions (round off) & 482 & & 142422 \\ 
Positions (x,y) & 514 & & 461873 \\
Double stars & 198 & 2063 & 1631 \\
\\ 
{\bf Photometry} \\
$UBV$ (photographic) & 294 & 126775 & 100221 \\ 
$UBV$ (photoelectric) & 439 & 34000 & 23038 \\ 
$UBV$ (CCD) & 261 & 315374 & 261135 \\ 
$VRI$ (Cousins) & 43 & 1460 & 412 \\
$VRI$ (Cousins; CCD) & 86 & 45596 & 42788 \\ 
$RI$ (Cousins; CCD) & 12 & 2803 & 2712 \\
$VI$ (Cousins; CCD) & 192 & 286357 & 257731 \\ 
$VRI$ (Johnson) & 97 & 2598 & 2145 \\ 
$uvby$ (photoelectric) & 214 & 7260 & 4949 \\ 
$uvby$ (CCD) & 25 & 21371 & 20277 \\ 
$\beta$ (photoelectric) & 248 & 7277 & 4771 \\ 
$\beta$ (CCD) & 16 & 2685 & 2414 \\ 
Geneva 7-color & 190 & 4618 & 4496 \\ 
$RGU$ (photographic) & 79 & 10369 & 10332 \\ 
\\
{\bf Spectroscopy} \\
MK types & 300 & 10397 & 6399 \\ 
HD types & 319 & 13148 & 12625 \\
$v$\,sin\,$i$ & 107 & 4636 & 3199 \\ 
$V_r$ (mean) & 92 & 3734 & 3492 \\ 
$V_r$ (individual) & 214 & 44606 & 5927 \\
$V_r$ (GPO) & 10 & 702 & 699 \\
$V_r$ (RFS) & 7 & 141 & 141 \\ 
Orbits & 59 & 419 & 275 \\ 
\\
{\bf Miscellaneous} \\
Proper motion (abs) & 7 & & 3653 \\ 
Proper motion (rel) & 12 & 6304 & 6302 \\ 
Probability ($\mu$) & 81 & & 39384 \\ 
Probability ($V_r$) & 8 & & 655 \\ 
Periods (Var) & 50 & 2482 & 1905 \\ 
X-ray flux & 28 & 3910 & 3351 \\ 
gK stars & 260 & & 5189 \\ 
Am stars & 34 & & 110 \\ 
Ap stars & 84 & & 218 \\ 
Be stars & 86 & & 368 \\ 
Blue stragglers & 209 & & 930 \\ 
Spectroscopic binaries & 49 & & 934 \\ 
\hline
\end{tabular}
\end{center}
\end{table}

\section{Description of the database}

WEBDA (accessible via http://obswww.unige.ch/webda/) and its predecessor BDA 
has been developed since 
1987 at the Institute for Astronomy, University of Lausanne, Switzerland by JCM. 
The progress of its development was described by Mermilliod (1988, 1992, 1995).
We will give here a brief overview of its current status and content. 

The database tries to collect all published data for stars in open clusters that 
may be useful either to determine the star membership, or to study the stellar 
content and properties of the clusters.

It is divided into three levels: 1) database; 2) cluster and
3) star level.

The database contents includes measurements in most photometric systems in which cluster 
stars have been observed, spectroscopic observations, astrometric data, various kinds of 
useful information, and extensive bibliography. It is possible to perform selection of 
clusters based on the amount of available data.
The data are usually recorded in their original form, with an indication of the source, 
but also as averaged values or selected data when relevant. 

The greatest effort has been spent in solving the identification problems raised by the 
definition of so many different numbering systems and a special interface has been developed 
to query the cross-reference tables.

Maps for more than 200 clusters have been scanned and included in the database. They are 
active maps and permit to retrieve basic data (e.g. positions, cross identifications 
and Johnson $UBV$ values) simply by clicking on the star images.

The database structure uses the directory hierarchy supported by the Unix system. 
The main directory is the database itself. It contains several sub-directories: 
description of the database, help information, references, bibliography, programs and 
perl scripts.
Each cluster defines an independent directory identified by its name and containing the 
available data in distinct files, one for each data type. This structure allows easy 
inclusion of any new cluster and any additional data type.

Whenever possible, the records of the various data files have the same structure: 
star identification, source, data. The files are organised sequentially and, within the 
files, the entries are sorted by star number and source reference.

The star identification is the main key to access the data, but it is also possible to use
filters based on the bibliographic references or astrophysical parameters.

The database engine WEBDA is a relational database built upon the package ``/rdb'' developed by 
Manis et al. (1988) which is a high performance relational database management and application 
development system designed for Unix environments.

Samples of clusters can be obtained by performing a selection on the clusters parameters, i.e. 
coordinates, distances, ages, diameters. The form prepared permits to do the selection on all 
parameters simultaneously. Clusters may also be selected on active plots, drawing the clusters 
in right ascension versus declination plane, galactic longitude versus latitude, 
distance from the Sun or above the galactic plane.

The database is in a dynamic growing process as new data are published and included. Table
\ref{content} lists an excerpt of the content of WEBDA based on the status from the 8th
April 2003. This date is also used as a ``deadline'' for our final analysis.

\begin{table*}
\begin{center}
\caption{573 open clusters with the number of objects 
from WEBDA with photometric measurements used for our analysis.}
\label{clusters}
{\tiny
\begin{tabular}{lr|lr|lr|lr|lr|lr|lr}
\hline
Cluster & N & Cluster & N & Cluster & N & Cluster & N & Cluster & N & Cluster & N & Cluster & N \\
\hline
Afgl 4029	&	8	&	Cr 347	&	20	&	Mel 71	&	795	&	NGC 2343	&	56	&	NGC 5823	&	163	&	NGC 7380	&	1686	&	Tr 17	&	147	\\
Am 2	&	2277	&	Cr 359	&	13	&	Mel 101	&	22	&	NGC 2345	&	64	&	NGC 5999	&	341	&	NGC 7419	&	716	&	Tr 18	&	153	\\
Bas 1	&	151	&	Cr 394	&	162	&	Mel 105	&	384	&	NGC 2353	&	141	&	NGC 6005	&	727	&	NGC 7510	&	674	&	Tr 21	&	368	\\
Bas 11a	&	89	&	Cr 399	&	8	&	Mel 111	&	435	&	NGC 2354	&	299	&	NGC 6025	&	179	&	NGC 7654	&	1247	&	Tr 22	&	100	\\
Bas 12	&	61	&	Cr 463	&	82	&	Mel 227	&	25	&	NGC 2355	&	829	&	NGC 6031	&	288	&	NGC 7686	&	81	&	Tr 24	&	442	\\
Bas 13	&	73	&	Cr 469	&	82	&	NGC 103	&	2836	&	NGC 2360	&	181	&	NGC 6067	&	1401	&	NGC 7762	&	580	&	Tr 26	&	98	\\
Bas 14	&	94	&	Cz 2	&	2351	&	NGC 129	&	1404	&	NGC 2362	&	100	&	NGC 6087	&	1334	&	NGC 7772	&	52	&	Tr 27	&	82	\\
Bas 15	&	107	&	Cz 8	&	19	&	NGC 133	&	312	&	NGC 2367	&	18	&	NGC 6124	&	299	&	NGC 7788	&	133	&	Tr 28	&	85	\\
Be 1	&	181	&	Cz 13	&	56	&	NGC 146	&	641	&	NGC 2374	&	83	&	NGC 6134	&	637	&	NGC 7789	&	16000	&	Tr 31	&	79	\\
Be 2	&	223	&	Cz 29	&	18	&	NGC 188	&	3893	&	NGC 2383	&	722	&	NGC 6167	&	48	&	NGC 7790	&	2470	&	Tr 32	&	1786	\\
Be 7	&	722	&	Do 24	&	8	&	NGC 189	&	93	&	NGC 2384	&	335	&	NGC 6178	&	58	&	NGC 7822	&	21	&	Tr 33	&	74	\\
Be 11	&	590	&	Do 25	&	128	&	NGC 225	&	326	&	NGC 2395	&	53	&	NGC 6192	&	242	&	Pis 1	&	23	&	Tr 35	&	306	\\
Be 12	&	1671	&	Do 42	&	37	&	NGC 366	&	1014	&	NGC 2414	&	12	&	NGC 6193	&	635	&	Pis 2	&	3536	&	Tr 37	&	291	\\
Be 14	&	1904	&	Eso92sc18	&	1804	&	NGC 381	&	2918	&	NGC 2420	&	910	&	NGC 6200	&	15	&	Pis 3	&	761	&	Tu 1	&	91	\\
Be 17	&	4050	&	Eso93sc08	&	1240	&	NGC 433	&	2119	&	NGC 2421	&	117	&	NGC 6204	&	160	&	Pis 4	&	16	&	Up 1	&	7	\\
Be 18	&	8734	&	Eso96sc04	&	999	&	NGC 436	&	897	&	NGC 2422	&	131	&	NGC 6208	&	243	&	Pis 5	&	9	&	Vdb 1	&	196	\\
Be 19	&	158	&	Ha 8	&	23	&	NGC 457	&	3888	&	NGC 2423	&	149	&	NGC 6216	&	199	&	Pis 8	&	26	&	Wat 3	&	7	\\
Be 20	&	429	&	Ha 20	&	28	&	NGC 559	&	217	&	NGC 2437	&	295	&	NGC 6231	&	1544	&	Pis 11	&	17	&	Wat 6	&	30	\\
Be 21	&	1645	&	Haf 6	&	699	&	NGC 581	&	5814	&	NGC 2439	&	305	&	NGC 6242	&	138	&	Pis 12	&	17	&	Wat 7	&	7	\\
Be 22	&	2017	&	Haf 8	&	78	&	NGC 609	&	84	&	NGC 2447	&	104	&	NGC 6249	&	15	&	Pis 16	&	115	&	Wes 1	&	233	\\
Be 23	&	1410	&	Haf 10	&	9	&	NGC 637	&	651	&	NGC 2451	&	322	&	NGC 6250	&	37	&	Pis 17	&	9	&	Wes 2	&	93	\\
Be 28	&	542	&	Haf 14	&	25	&	NGC 654	&	666	&	NGC 2451A	&	136	&	NGC 6253	&	7975	&	Pis 18	&	344	&		&		\\
Be 29	&	1125	&	Haf 15	&	13	&	NGC 659	&	767	&	NGC 2451B	&	19	&	NGC 6259	&	563	&	Pis 19	&	5183	&		&		\\
Be 30	&	1923	&	Haf 16	&	15	&	NGC 663	&	3765	&	NGC 2453	&	382	&	NGC 6268	&	75	&	Pis 20	&	219	&		&		\\
Be 31	&	2075	&	Haf 17	&	122	&	NGC 744	&	117	&	NGC 2467	&	352	&	NGC 6281	&	85	&	Pis 21	&	294	&		&		\\
Be 32	&	3283	&	Haf 18	&	78	&	NGC 752	&	589	&	NGC 2477	&	19384	&	NGC 6318	&	244	&	Pis 22	&	198	&		&		\\
Be 33	&	1869	&	Haf 19	&	280	&	NGC 869	&	3816	&	NGC 2482	&	41	&	NGC 6322	&	113	&	Pis 23	&	627	&		&		\\
Be 39	&	4395	&	Haf 20	&	33	&	NGC 884	&	3300	&	NGC 2483	&	75	&	NGC 6383	&	595	&	Pis 24	&	17	&		&		\\
Be 42	&	556	&	Haf 21	&	51	&	NGC 957	&	255	&	NGC 2489	&	155	&	NGC 6396	&	22	&	Pl 1	&	152	&		&		\\
Be 54	&	2495	&	Her 1	&	16	&	NGC 1027	&	153	&	NGC 2506	&	1417	&	NGC 6405	&	635	&	Ros 3	&	83	&		&		\\
Be 58	&	420	&	Hm 1	&	803	&	NGC 1039	&	1078	&	NGC 2516	&	2558	&	NGC 6416	&	330	&	Ros 4	&	14	&		&		\\
Be 60	&	2121	&	Ho 9	&	9	&	NGC 1193	&	503	&	NGC 2527	&	404	&	NGC 6425	&	74	&	Ros 5	&	46	&		&		\\
Be 62	&	1583	&	Ho 10	&	24	&	NGC 1220	&	234	&	NGC 2533	&	124	&	NGC 6451	&	744	&	Ru 18	&	20	&		&		\\
Be 64	&	2042	&	Ho 11	&	6	&	NGC 1245	&	712	&	NGC 2539	&	354	&	NGC 6475	&	896	&	Ru 20	&	11	&		&		\\
Be 65	&	42	&	Ho 12	&	11	&	NGC 1252	&	41	&	NGC 2546	&	688	&	NGC 6494	&	218	&	Ru 32	&	133	&		&		\\
Be 66	&	1677	&	Ho 14	&	11	&	NGC 1342	&	311	&	NGC 2547	&	227	&	NGC 6514	&	311	&	Ru 34	&	17	&		&		\\
Be 68	&	126	&	Ho 15	&	454	&	NGC 1348	&	1030	&	NGC 2548	&	47	&	NGC 6520	&	412	&	Ru 36	&	72	&		&		\\
Be 69	&	144	&	Ho 16	&	86	&	NGC 1444	&	99	&	NGC 2567	&	275	&	NGC 6530	&	1028	&	Ru 44	&	82	&		&		\\
Be 70	&	2464	&	Ho 17	&	41	&	NGC 1496	&	51	&	NGC 2571	&	1662	&	NGC 6531	&	408	&	Ru 46	&	597	&		&		\\
Be 79	&	60	&	Ho 18	&	28	&	NGC 1502	&	155	&	NGC 2579	&	56	&	NGC 6546	&	52	&	Ru 47	&	10	&		&		\\
Be 81	&	3301	&	Ho 22	&	30	&	NGC 1513	&	228	&	NGC 2627	&	507	&	NGC 6603	&	3598	&	Ru 49	&	9	&		&		\\
Be 82	&	20	&	IC 166	&	208	&	NGC 1528	&	619	&	NGC 2632	&	605	&	NGC 6604	&	117	&	Ru 55	&	29	&		&		\\
Be 86	&	736	&	IC 348	&	201	&	NGC 1545	&	67	&	NGC 2635	&	6	&	NGC 6611	&	1359	&	Ru 59	&	21	&		&		\\
Be 87	&	105	&	IC 361	&	19	&	NGC 1605	&	38	&	NGC 2645	&	74	&	NGC 6613	&	119	&	Ru 67	&	27	&		&		\\
Be 93	&	87	&	IC 1311	&	976	&	NGC 1624	&	14	&	NGC 2658	&	123	&	NGC 6618	&	671	&	Ru 76	&	7	&		&		\\
Be 94	&	50	&	IC 1369	&	155	&	NGC 1647	&	362	&	NGC 2659	&	16	&	NGC 6631	&	5533	&	Ru 79	&	361	&		&		\\
Be 96	&	10	&	IC 1442	&	105	&	NGC 1662	&	73	&	NGC 2660	&	914	&	NGC 6633	&	693	&	Ru 82	&	144	&		&		\\
Be 99	&	867	&	IC 1590	&	255	&	NGC 1664	&	318	&	NGC 2669	&	31	&	NGC 6649	&	566	&	Ru 83	&	93	&		&		\\
Be 104	&	3173	&	IC 1795	&	191	&	NGC 1750	&	7396	&	NGC 2670	&	393	&	NGC 6664	&	60	&	Ru 92	&	59	&		&		\\
Bh 66	&	735	&	IC 1805	&	1984	&	NGC 1778	&	140	&	NGC 2671	&	62	&	NGC 6683	&	163	&	Ru 93	&	93	&		&		\\
Bh 99	&	621	&	IC 1848	&	74	&	NGC 1798	&	1416	&	NGC 2682	&	3192	&	NGC 6694	&	122	&	Ru 97	&	251	&		&		\\
Bh 176	&	9999	&	IC 2157	&	2017	&	NGC 1807	&	39	&	NGC 2818	&	624	&	NGC 6704	&	569	&	Ru 98	&	16	&		&		\\
Bh 222	&	301	&	IC 2391	&	329	&	NGC 1817	&	370	&	NGC 2866	&	23	&	NGC 6705	&	8377	&	Ru 103	&	163	&		&		\\
Bh 245	&	122	&	IC 2395	&	61	&	NGC 1857	&	79	&	NGC 2910	&	134	&	NGC 6709	&	1365	&	Ru 107	&	17	&		&		\\
Biu 2	&	132	&	IC 2488	&	145	&	NGC 1893	&	1656	&	NGC 2925	&	185	&	NGC 6716	&	888	&	Ru 108	&	11	&		&		\\
Bl 1	&	355	&	IC 2581	&	398	&	NGC 1901	&	43	&	NGC 2972	&	14	&	NGC 6755	&	310	&	Ru 115	&	486	&		&		\\
Bo 1	&	15	&	IC 2602	&	376	&	NGC 1907	&	324	&	NGC 3033	&	19	&	NGC 6756	&	402	&	Ru 118	&	7	&		&		\\
Bo 2	&	87	&	IC 2714	&	224	&	NGC 1912	&	778	&	NGC 3105	&	131	&	NGC 6791	&	9229	&	Ru 119	&	239	&		&		\\
Bo 3	&	8	&	IC 2944	&	138	&	NGC 1931	&	163	&	NGC 3114	&	2277	&	NGC 6802	&	225	&	Ru 120	&	149	&		&		\\
Bo 4	&	30	&	IC 4651	&	15845	&	NGC 1960	&	1132	&	NGC 3228	&	434	&	NGC 6811	&	1018	&	Ru 124	&	424	&		&		\\
Bo 6	&	5	&	IC 4665	&	429	&	NGC 1976	&	3192	&	NGC 3255	&	8	&	NGC 6819	&	2565	&	Ru 127	&	18	&		&		\\
Bo 7	&	1433	&	IC 4725	&	1461	&	NGC 2099	&	3896	&	NGC 3293	&	511	&	NGC 6823	&	890	&	Ru 129	&	55	&		&		\\
Bo 8	&	8	&	IC 4756	&	507	&	NGC 2112	&	612	&	NGC 3324	&	988	&	NGC 6830	&	158	&	Ru 130	&	345	&		&		\\
Bo 9	&	2907	&	IC 4996	&	718	&	NGC 2129	&	203	&	NGC 3330	&	66	&	NGC 6834	&	1251	&	Ru 140	&	259	&		&		\\
Bo 10	&	425	&	IC 5146	&	734	&	NGC 2141	&	3309	&	NGC 3496	&	272	&	NGC 6866	&	599	&	Ru 146	&	163	&		&		\\
Bo 11	&	514	&	Ki 2	&	1031	&	NGC 2158	&	4672	&	NGC 3532	&	728	&	NGC 6871	&	1979	&	Ru 166	&	954	&		&		\\
Bo 12	&	12	&	Ki 4	&	151	&	NGC 2168	&	2102	&	NGC 3572	&	85	&	NGC 6882	&	76	&	Ru 175	&	113	&		&		\\
Bo 13	&	13	&	Ki 5	&	1347	&	NGC 2169	&	36	&	NGC 3590	&	79	&	NGC 6883	&	196	&	Sh 1	&	41	&		&		\\
Bo 14	&	11	&	Ki 6	&	475	&	NGC 2175	&	155	&	NGC 3603	&	515	&	NGC 6910	&	234	&	Sha 138	&	259	&		&		\\
Bo 15	&	33	&	Ki 7	&	698	&	NGC 2186	&	23	&	NGC 3680	&	905	&	NGC 6913	&	464	&	St 1	&	160	&		&		\\
Cr 69	&	132	&	Ki 8	&	259	&	NGC 2192	&	409	&	NGC 3766	&	2658	&	NGC 6939	&	462	&	St 2	&	4297	&		&		\\
Cr 74	&	739	&	Ki 9	&	2058	&	NGC 2194	&	2146	&	NGC 3960	&	317	&	NGC 6940	&	395	&	St 7	&	29	&		&		\\
Cr 96	&	14	&	Ki 10	&	1183	&	NGC 2204	&	2771	&	NGC 4103	&	4091	&	NGC 6994	&	197	&	St 8	&	23	&		&		\\
Cr 97	&	29	&	Ki 11	&	1163	&	NGC 2215	&	43	&	NGC 4337	&	18	&	NGC 7031	&	73	&	St 13	&	112	&		&		\\
Cr 107	&	267	&	Ki 12	&	31	&	NGC 2232	&	43	&	NGC 4349	&	216	&	NGC 7039	&	220	&	St 14	&	137	&		&		\\
Cr 110	&	471	&	Ki 13	&	80	&	NGC 2236	&	495	&	NGC 4439	&	24	&	NGC 7044	&	2531	&	St 16	&	104	&		&		\\
Cr 121	&	47	&	Ki 14	&	196	&	NGC 2243	&	3705	&	NGC 4463	&	20	&	NGC 7062	&	431	&	St 17	&	10	&		&		\\
Cr 132	&	35	&	Ki 15	&	2771	&	NGC 2244	&	1253	&	NGC 4609	&	52	&	NGC 7063	&	103	&	St 24	&	2121	&		&		\\
Cr 135	&	77	&	Ki 19	&	264	&	NGC 2251	&	615	&	NGC 4755	&	8612	&	NGC 7067	&	85	&	Ste 1	&	179	&		&		\\
Cr 140	&	80	&	Ki 21	&	26	&	NGC 2254	&	97	&	NGC 4815	&	8596	&	NGC 7082	&	182	&	Ter 7	&	1731	&		&		\\
Cr 185	&	74	&	Ly 1	&	24	&	NGC 2259	&	1422	&	NGC 5138	&	92	&	NGC 7086	&	220	&	To 1	&	1000	&		&		\\
Cr 197	&	21	&	Ly 2	&	97	&	NGC 2264	&	1791	&	NGC 5168	&	307	&	NGC 7092	&	193	&	To 2	&	2905	&		&		\\
Cr 205	&	18	&	Ly 4	&	6	&	NGC 2266	&	464	&	NGC 5281	&	1434	&	NGC 7127	&	70	&	Tr 1	&	1431	&		&		\\
Cr 223	&	110	&	Ly 6	&	124	&	NGC 2269	&	12	&	NGC 5316	&	131	&	NGC 7128	&	513	&	Tr 2	&	129	&		&		\\
Cr 228	&	1193	&	Ly 7	&	19	&	NGC 2281	&	1113	&	NGC 5367	&	10	&	NGC 7142	&	520	&	Tr 5	&	5150	&		&		\\
Cr 232	&	122	&	Ly 14	&	16	&	NGC 2287	&	217	&	NGC 5381	&	3239	&	NGC 7160	&	341	&	Tr 7	&	16	&		&		\\
Cr 258	&	36	&	Ma 38	&	36	&	NGC 2301	&	1608	&	NGC 5460	&	328	&	NGC 7209	&	119	&	Tr 9	&	52	&		&		\\
Cr 261	&	3523	&	Ma 50	&	256	&	NGC 2302	&	16	&	NGC 5606	&	191	&	NGC 7226	&	259	&	Tr 10	&	57	&		&		\\
Cr 268	&	23	&	Mel 20	&	701	&	NGC 2304	&	1449	&	NGC 5617	&	468	&	NGC 7235	&	666	&	Tr 11	&	355	&		&		\\
Cr 271	&	10	&	Mel 22	&	770	&	NGC 2323	&	253	&	NGC 5662	&	910	&	NGC 7243	&	60	&	Tr 14	&	586	&		&		\\
Cr 272	&	1249	&	Mel 25	&	1430	&	NGC 2324	&	213	&	NGC 5749	&	112	&	NGC 7245	&	338	&	Tr 15	&	869	&		&		\\
Cr 307	&	12	&	Mel 66	&	3909	&	NGC 2335	&	63	&	NGC 5822	&	709	&	NGC 7261	&	148	&	Tr 16	&	461	&		&		\\
\hline
\end{tabular}
}
\end{center}
\end{table*}

\begin{table*}[t]
\begin{center}
\caption{Deviating data sets within one photometric system. The errors in the final 
digits of the corresponding quantity are given in parenthesis.}
\label{within}
\begin{tabular}{llcclr}
\hline
Cluster & Photometric system & Set1 & Set2 & \multicolumn{1}{c}{mean} & N(obj) \\
\hline
Berkeley 64 & $UBV$: $V$ (CCD) & Pandey et al. (1995) & Ann et al. (2002) & 
$-$0.161(44) & 26 \\
Markarian 50 & $UBV$: $U-B$ (pgo) & Turner et al. (1983) & Grubissich (1965) & 
+0.174(27) & 27 \\
Melotte 71 & $UBV$: $V$ (pgo) & Pound \& Janes (1986) & Hassan (1976) & +0.388(102) & 33 \\
NGC 1348 & $VIc$: $V$ & Ann et al. (2002) & Carraro (2002) & +0.197(55) & 124 \\
NGC 2244 & $VIc$: $V-I$ & Park \& Sung (2002) & Bergh{\"o}fer \& Christian (2002) 
& $-$0.237(68) & 124 \\
NGC 6611 & $UBV$: $U-B$ (peo) & Hiltner \& Morgan (1969) & Th{\'e} et al. (1989) & 
+0.083(22) & 15 \\
NGC 6791 & $VIc$: $V-I$ & Garnavich et al. (1994) & von Braun et al. (1998) & +0.104(34) & 12 \\
NGC 6910 & $UBV$: $V$ (peo) & Hoag et al. (1961) & Heiser (private communication) & 
+0.088(26) & 13 \\
NGC 7044 & $UBV$: $V$ (CCD) & Aparicio et al. (1993) & Sagar \& Griffiths (1998) & 
+0.135(36) & 553 \\
NGC 7654 & $UBV$: $V$ (CCD) & Choi et al. (1999) & Stetson (2000) & $-$0.164(23) & 49 \\
		 & $UBV$: $B-V$ (CCD) & Choi et al. (1999) & Stetson (2000) & +0.153(22) & 73 \\
		 & $VIc$: $V-I$ & Choi et al. (1999) & Stetson (2000) & $-$0.231(32) & 57 \\
\hline
\end{tabular}
\end{center}
\end{table*}

\section{Analysis and compilation of the data} \label{analysis}

The final goal of this extensive statistical analysis is not only
to investigate the consistency of the published data but more importantly
to derive ages, reddening values and distances for a large number
of open clusters. It is therefore necessary to select photometric systems
for which enough data and appropriate isochrones are available. Table \ref{content}
lists the numbers for the most common photometric systems included in WEBDA.
From a close inspection we have chosen the following photometric systems
for our analysis:
\begin{itemize}
\item Johnson $UBV$; photographic, photoelectric and CCD measurements
\item Cousins $VI$; CCD
\item Cousins $VRI$; photoelectric and CCD
\item Str{\"o}mgren $uvby\beta$; photoelectric and CCD
\item $RGU$; photographic, for comparison
\end{itemize}
We have not included the Johnson $VI$ systems because there
are usually only a few measurements for the brightest objects within one cluster
making an isochrone fitting impossible. The Geneva 7-color system is
outstanding compared to the other photometric systems. For most of the open clusters,
only the brightest members are investigated whereas for
a few ones also the faintest members were observed. Furthermore, WEBDA already includes
the mean values for all these objects, so no improvement can be done within
our analysis. We have used
the data of the Geneva 7-color system for several open clusters in order to check the results
from the other photometric systems.

The list with available
photometric data in the mentioned above systems contains 573 open clusters
(listed in Table \ref{clusters}) with a total of 469820 objects.

The data analysis of the relevant photometric systems includes several
different steps in order to perform a careful check of the homogeneity
of the individual sources. Much effort was already spent to improve the
homogeneity of WEBDA by investigating the published data, finding charts and listed
coordinates (Mermilliod 1988, 1992, 1995). This process is very time consuming and
not straightforward. During the first stages of our new analysis
we have already found some wrongly identified objects and misprints in the
literature. These errors have already been fixed in WEBDA. But we have to
emphasize that these are only the ``eye hitting'' divergences, still there are 
many unsolved cases (see the lists mentioned above) which have to be investigated
in the future.

As the first step of our analysis we have checked the intrinsic consistency
of different sources for one photometric system (e.g. photographic Johnson
$UBV$ data) of all individual open clusters. In general, we have used the following
(very conventional) limits for a measurement being ``oustanding'' if the difference
of the data are larger than:
\begin{itemize}
\item 0.5\,mag: $UBV$ photographic; $RGU$
\item 0.2\,mag: $UBV$, $VRIc$, photoelectric and CCD; $uvby$, photoelectric
\item 0.1\,mag: $uvby$, CCD; $\beta$ photoelectric and CCD
\end{itemize}
The compiled list includes 4467 entries (2914 from photographic measurements). 
Excluding these objects, we have searched for intrinsic 
correlations for data sets which have more than five objects in common using a simple
linear correlation algorithm. We only
find twelve statistically significant deviating cases for ten open clusters. These
deviating cases are listed in Table \ref{within}. Paunzen \& Maitzen (2002) reported
one deviating case for NGC 6451 for which they were able to show that the
photometry by Piatti et al. (1998) has an unidentified error and was therefore excluded
from our analysis.

\begin{figure}
\begin{center}
\includegraphics[width=80mm]{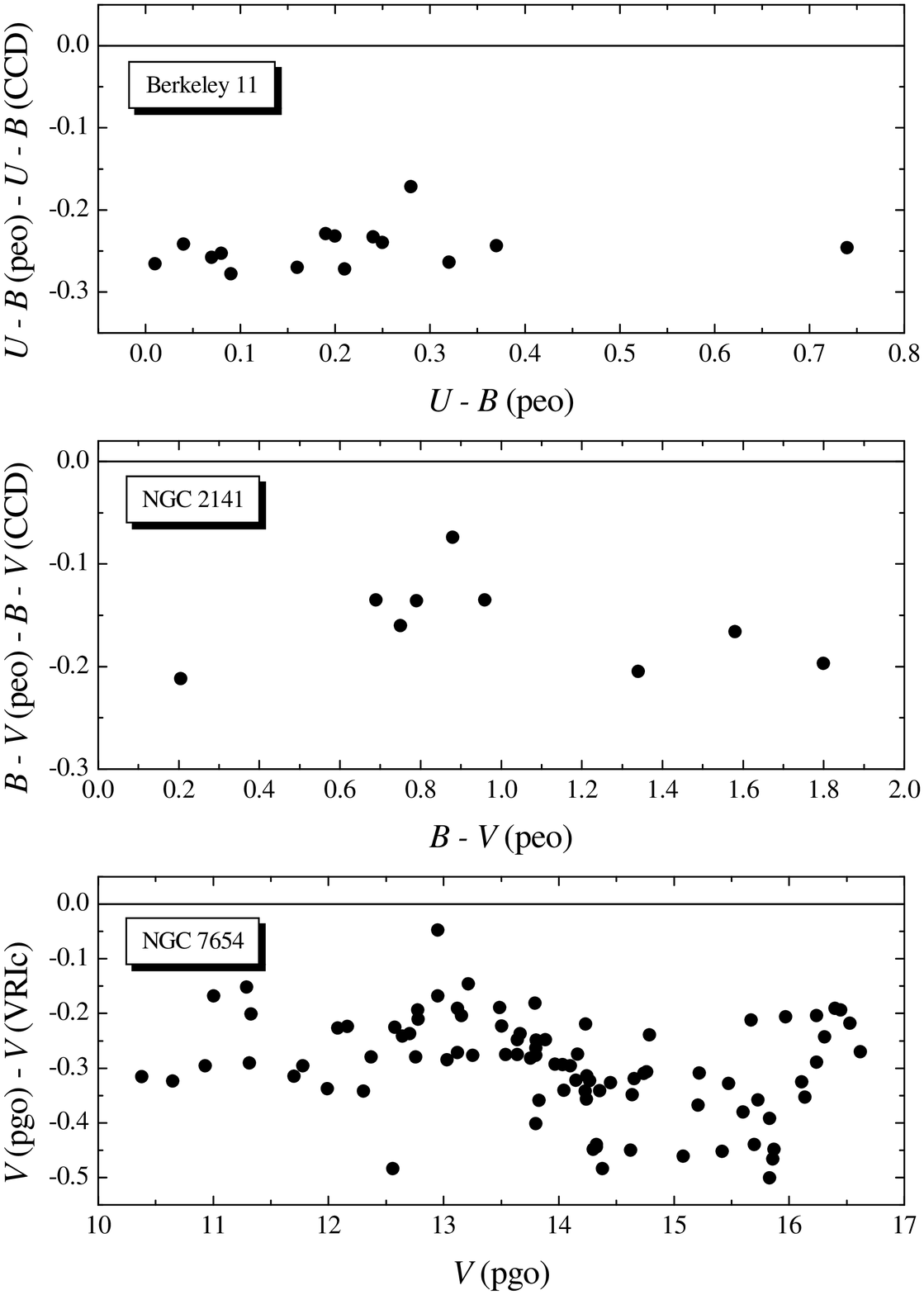}
\caption{Three examples for deviating measurements from different
observing techniques as listed in Table \ref{without}.}
\label{fig_ext}
\end{center}
\end{figure}

As a next step we have used the averaged mean values of different 
photometric systems (e.g. $UBV$ photographic and photoelectric) to search
for ``external'' discrepancies between measurements for the individual
clusters. Again, a list of outstanding objects was created with the limits
given as:
\begin{equation}
limit(ext) = \sqrt{limit(system_1)^2+limit(system_2)^2}
\end{equation}
This list has 7061 entries. Table \ref{without} shows the deviating
data sets from this external check. Figure \ref{fig_ext} shows three
examples for the Johnson $UBV$ system graphically.

Since it is well known that photographic
measurements have in general larger errors we have used averages of photoelectric
and CCD data only. If such data are not available, photographic ones were used.
The final averaged weighted values were calculated following the approach
described in Mermilliod \& Mermilliod (1994) which is a two step iterative procedure.
The first step consists of a weighted mean, the weight being the number 
of measurements to the 2/3 power. The next step uses the differences between the
weighted mean and the individual values to compute new weighted mean values. This
procedure gives a lower weight to discrepant values.

We also find a few cases (e.g. for NGC 6705) for which data sets show a trend
within different photometric systems, e.g. $(U-B)_{Set1} - (U-B)_{Set2}$ versus
$(B-V)_{Set1}$. There is no straightforward solution for these data sets. However,
we have excluded those data from our final analysis.

The complete tables with the available weighted mean values will be available at
WEBDA only or upon request from the authors.

\begin{table}[t]
\begin{center}
\caption{Deviating data sets within different photometric systems.}
\label{without}
\begin{tabular}{llrrrr}
\hline
Cluster & Photometric system & \multicolumn{1}{c}{mean} & N(obj) \\
\hline
Berkeley 11 & $U-B$ (peo, CCD) & $-$0.247(25) & 15 \\
		    & $B-V$ (peo, CCD) & $-$0.083(25) & 17 \\
Berkeley 58 & $V$ (pgo, CCD) & $-$0.346(84) & 36 \\
Bochum 10 & $V$ (pgo, CCD) & $-$0.346(84) & 36 \\
Haffner 8 & $U-B$ (pgo, peo) & $-$0.303(86) & 7 \\
NGC 637 & $V$ (pgo, peo) & $-$0.268(60) & 25 \\
        & $V$ (pgo, CCD) & $-$0.303(76) & 35 \\
NGC 884 & $m_1$ (peo, CCD) & +0.130(35) & 17 \\
NGC 1245 & $V$ (peo, CCD) & $-$0.241(40) & 15 \\
NGC 2112 & $V$ (peo, CCD) & +0.168(46) & 16 \\
         & $V$ (peo, VRIc) & $-$0.241(40) & 15 \\
		 & $V$ (CCD, uvby) & +0.221(52) & 9 \\
NGC 2141 & $B-V$ (peo, CCD) & $-$0.158(41) & 9 \\
NGC 2158 & $V$ (peo, CCD) & $-$0.129(24) & 8 \\
         & $V$ (peo, VIc) & $-$0.124(32) & 8 \\
NGC 3680 & $V$ (CCD, VIc) & +0.077(12) & 6 \\
NGC 6005 & $V$ (CCD, VIc) & $-$0.053(17) & 529 \\
NGC 7654 & $V$ (pgo, VRIc) & $-$0.296(89) & 89 \\
         & $V$ (peo, VRIc) & $-$0.157(28) & 18 \\
         & $V$ (VRIc, uvby) & +0.147(41) & 9 \\
\hline
\end{tabular}
\end{center}
\end{table}

\begin{figure*}
\begin{center}
\includegraphics[width=140mm]{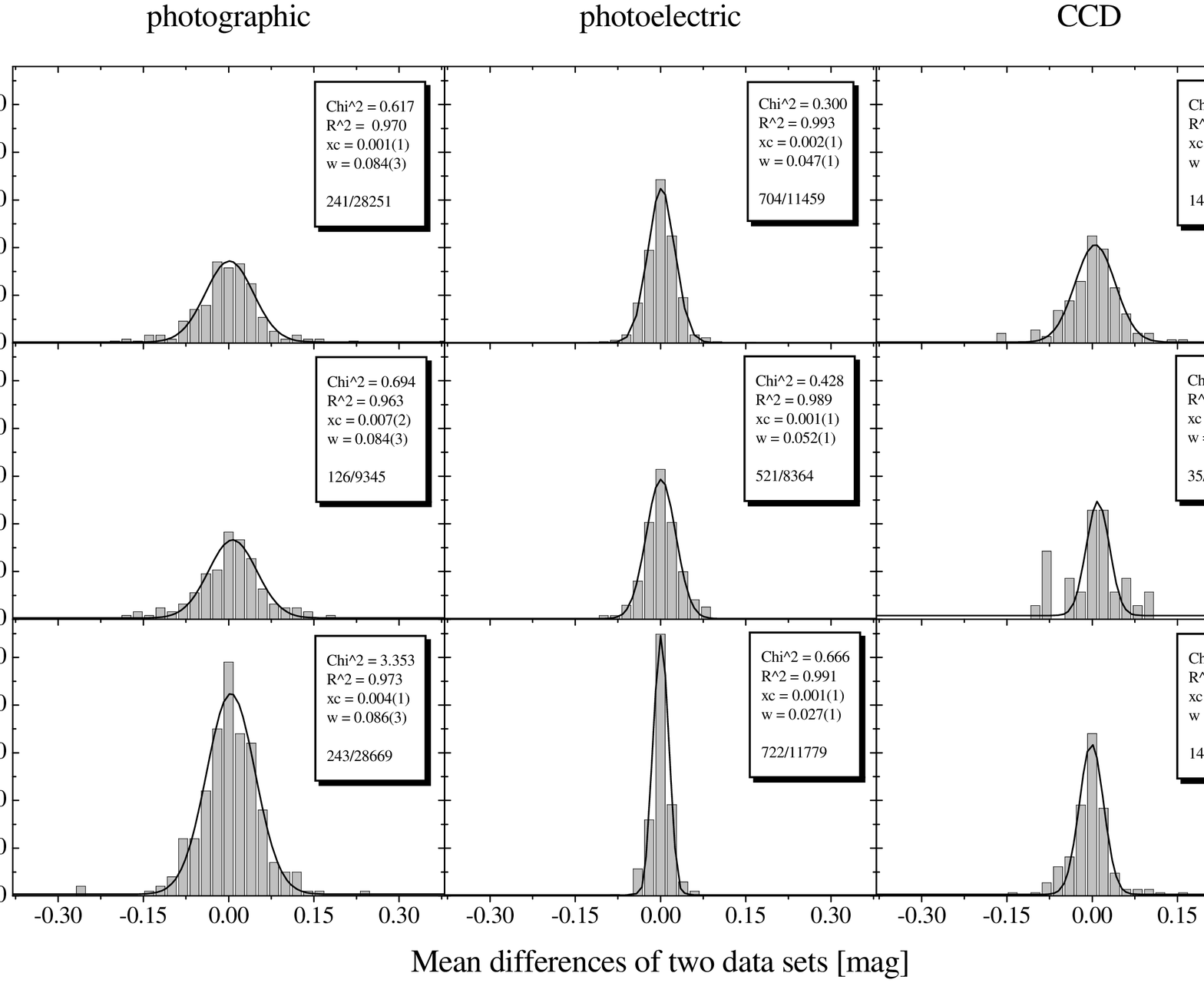}
\caption{Histograms of the internal accuracy of the Johnson $UBV$ photometric
system for photographic (left panel), photoelectric (middle panel) and CCD (left
panel) measurements; listed are the most important parameters of the fitted
Gaussian distributions together with the number of data sets and objects
($N_1/N_2$).}
\label{int_ubv}
\end{center}
\end{figure*}

\begin{figure}
\begin{center}
\includegraphics[width=88mm]{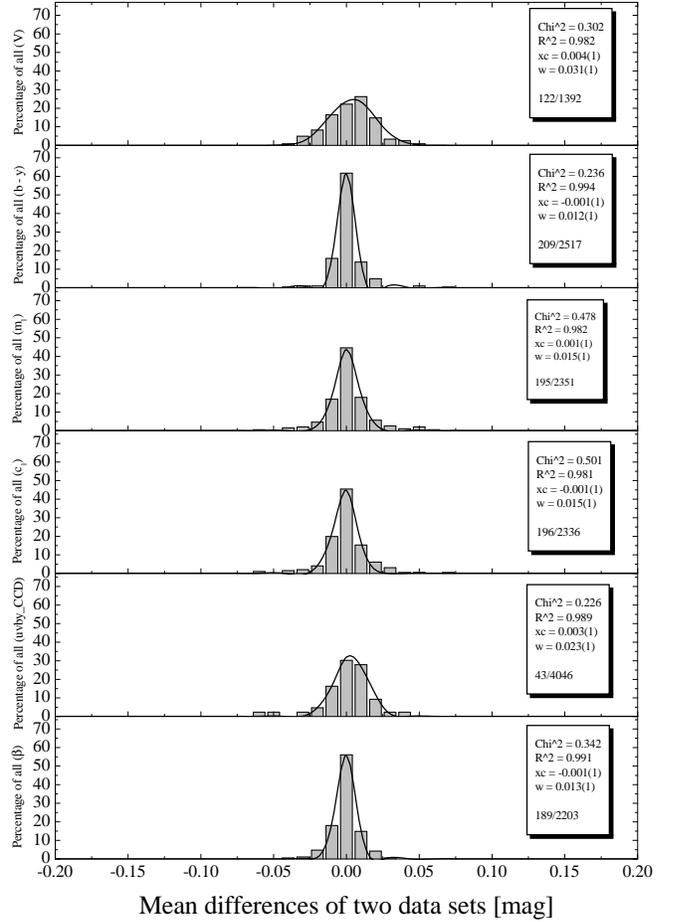}
\caption{Histograms of the internal accuracy of the Str{\"o}mgren $uvby\beta$ photometric
system. The upper four panels show the results for photoelectric $uvby$,
the fifth CCD $uvby$ and the last panel all $\beta$ measurements; listed are the most 
important parameters of the fitted
Gaussian distributions together with the number of data sets and objects
($N_1/N_2$).}
\label{int_uvby}
\end{center}
\end{figure}

\begin{figure}
\begin{center}
\includegraphics[width=88mm]{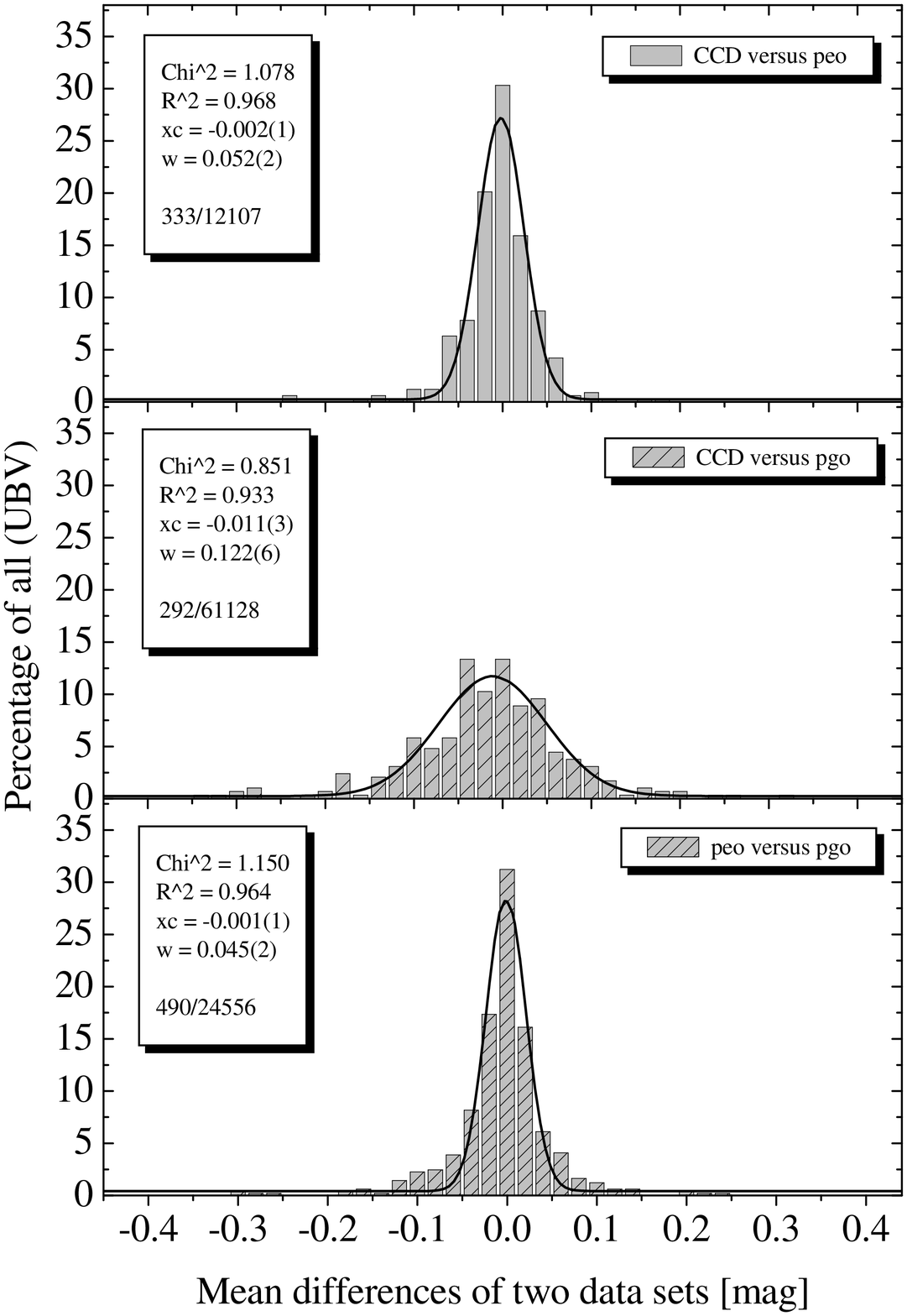}
\caption{Histograms of the external accuracy of the Johnson $UBV$ photometric
system for CCD versus photoelectric (upper panel) and photographic (middle panel) as
well as photoelectric versus photographic (lower panel) measurements; 
listed are the most important parameters of the fitted
Gaussian distributions together with the number of data sets and objects
($N_1/N_2$).}
\label{ext_ubv}
\end{center}
\end{figure}

\section{Discussion}

For the analysis of the overall accuracies of the available data sets, we have
used Gaussian distributions (Christensen 1996) to fit the histograms of all mean values for the
different data. The calculated histograms were normalized to the overall percentage
of the sum.
The results for the Johnson $UBV$ and Str{\"o}mgren $uvby\beta$
photometric systems are summarized in Figs \ref{int_ubv} to \ref{ext_ubv}. These
figures include the most important fit parameters such as the mean value, width,
$R^2$, $\chi^2$ as well as the the number of data sets and objects. 
The corresponding
histograms in the other photometric systems are not plotted since there are too few
data points to compare which makes a statistically sound analysis impossible.

The way of calibrating observed magnitudes is either to simultaneously measure
``well established'' standard stars or to use already known standard transformations
for the individual telescope and filter set. Both approaches are certainly not straightforward.
Sung \& Bessell (2000) summarize and discuss the problems concerning the variations
of atmospheric extinction coefficients, transformation equations, different filter
systems, CCDs as well as the difference between two sets of standard stars (SAAO and
Landolt). They also include a compelling list of references concerning this special
topic. 

It is out of the scope of this statistical analysis to reproduce the used transformation
technique of the individual references. We have to rely on the published data. The only
possible check is to search for misidentified objects or typos. Otherwise undetected
variability of any kind could also lead to several divergent observations, again a
fact which we are not able to prove.

We will now discuss the internal and the external accuracy separately. 

\subsection{The internal accuracy}

The most important check for the reliability of published data is the comparison
with other independent measurements within the same photometric system and the
same technique. Figure \ref{int_ubv} shows the histograms (bin size is 0.02\,mag)
for the Johnson $UBV$ system. We have separated the photographic, photoelectric and 
CCD measurements.

The histograms are based on a statistically
significant number of individual data sets and objects.
The only exception is CCD measurements for $U-B$. This is probably caused by the 
insensitivity of the modern CCD detectors in the ultraviolet region which makes the observations
in the standard $U$ band almost impossible. From Fig. \ref{int_ubv} we are able to conclude:
\begin{itemize}
\item The bandwidth of the Gaussian distributions for the photographic measurements
is twice ($\approx$0.09\,mag) as large as the corresponding ones from the other two sources
\item The only exception is the CCD $V$ data which might be due to the relative faintness
and thus the larger observational error reach with this technique
\item All mean values of the fitted distributions are close to zero
\end{itemize}

The corresponding histograms for the Str{\"o}mgren $uvby\beta$ photometric system have 
a bin size of 0.01\,mag and are shown in Fig \ref{int_uvby}. The widths of the fitted
distributions are all between 0.01 and 0.03\,mag which shows the high quality of the
published data.

In total, 4467 deviating measurements were found (Sect. \ref{analysis}) within 4056 
different data sets and 266779 objects. This corresponds to 1.7\,\% which is an extremely
low percentage. If we exclude the outlyers from photographic measurements, this percentage
even lowers to 0.6\,\%.

\subsection{The external accuracy}

After having shown that the accuracies within the individual photometric systems
are very good. We then investigated the errors for different measurement techniques
and thus mainly different quantum efficiencies characteristics of detectors. 

Figure \ref{ext_ubv} shows the result for the Johnson $UBV$ photometric system (bin
size is 0.02\,mag). The
results for the other systems are similar. We have summarized the data for $V$, $B-V$
and $U-B$, otherwise we would run into poor number statistics. However, it shows
that the distributions for the three different indices with the same detector
technique are essentially the same.

The bandwidth of the Gaussian distributions for comparison of the photoelectric data sets
(upper and lower panel) is about 0.05\,mag whereas the comparison of the CCD versus
photographic measurements results in an almost three times higher value (0.12\,mag).
This reflects the most different quantum efficiency characteristics of these detectors
whereas the photoelectric one is in between. In addition, the larger scatter may be
due to the faintness of the photographically observed objects. Usually, the photoelectric
data magnitude limit is brighter than that of the photographic one.
We conclude that the photoelectric
measurements are still the most valuable to connect the photographic to the CCD ones.

This analysis is based on 1960 data sets with 292770 measurements. Again, the number
of deviating data points is surprisingly low (7061 or 2.4\,\%). This shows the high capability
of WEBDA to analyse the astrophysical properties of open clusters in the Milky Way.

\section{Conclusions and outlook}

The enormous amount of photometric data within WEBDA was analysed in order to
check for the internal and external accuracies of published data for open clusters.
This analysis is based on photographic, photoelectric and CCD measurements for
five different photometric systems which includes 573 open clusters and 469820 objects.
The way of weighting and averaging the data is described.

We have investigated 4056 data sets which have more than five objects in common and concluded
that the internal accuracies are very good. The accuracy is best for the Str{\"o}mgren $uvby\beta$
system and drops significantly towards photographic Johnson $UBV$ data.
Less than 2\,\% of deviating measurements were found and tabulated. 

A surprisingly good agreement between photoelectric and photographic as well as CCD
data was found. The higher error for CCD versus photographic data reflects the differences
of the individual quantum efficiency curves of these systems. Nevertheless, the amount
and homogenity of data will allow us to derive astrophysical parameters such
as the ages, distances and reddenings for the 573 open clusters investigated. This will
be done in a second paper which includes isochrone fitting and the discussion of
different statistical issues concerning the structure of our Milky Way. 

\begin{acknowledgements}
EP acknowledges partial support by the Fonds zur F\"orderung der
wissenschaftlichen Forschung, project P14984.
\end{acknowledgements}

\end{document}